\documentclass[review]{elsarticle}

\usepackage{lineno,hyperref,xcolor,amsmath}

\journal{Advances in Space Research}

\biboptions{authoryear}

\begin{document}

\begin{frontmatter}

\title{Simulation of Cosmic Rays in the Earth's Atmosphere and Interpretation of
Observed Counts in an X-ray Detector at Balloon Altitude Near Tropical Region}


\author[address]{Ritabrata Sarkar\corref{mycorrespondingauthor}}
\ead{ritabrata.s@gmail.com}
\cortext[mycorrespondingauthor]{Corresponding author}

\author[address]{Abhijit Roy}
\ead{aviatphysics@gmail.com}

\author[address]{Sandip K. Chakrabarti}
\ead{sandipchakrabarti9@gmail.com}

\address[address]{Indian Centre for Space Physics, 43 Chalantika, Garia Station
Rd., Kolkata 700084, W.B., India}

\begin{abstract}
The study of secondary particles produced by the cosmic-ray interaction in the
Earth's atmosphere is very crucial as these particles mainly constitute the
background counts produced in the high-energy detectors at balloon and satellite
altitudes. In the present work, we calculate the abundance of cosmic-ray
generated secondary particles at various heights of the atmosphere by means of a
Monte Carlo simulation and use this result to understand the background counts
in our X-ray observations using balloon-borne instruments operating near the
tropical latitude (geomagnetic latitude: $\sim 14.50^{\circ}$ N). For this
purpose, we consider a 3D description of the atmospheric and geomagnetic field
configurations surrounding the Earth, as well as the electromagnetic and nuclear
interaction processes using Geant4 simulation toolkit. Subsequently, we use a
realistic mass model description of the detector under consideration, to
simulate the counts produced in the detector due to secondary cosmic-ray
particles.
\end{abstract}


\end{frontmatter}


\section{Introduction} 
\label{sec:intro}
Galactic Cosmic Rays (GCRs) are originated from various cosmic sources outside
the solar system. They are primarily composed of protons (H $\sim 89\%$), alpha
particles (He $\sim 10\%$) and other heavier nuclei. On arriving the Earth's
atmosphere they interact with atmospheric nuclei to produce several secondary
leptonic and hadronic particles and radiation in the form of a shower. However,
before reaching the top of the atmosphere these GCR particles face dynamic
electromagnetic environment in the heliosphere which varies with solar activity 
as well as the geomagnetic field which acts as a shield.

The Coronal Mass Ejection (CME) and solar flares during a period of solar activity
affect the space environment in more than one way. A solar activity modulates the
flux of the primary GCRs, particularly the low energy part below a few GeV. The
Solar Energetic Particles (SEPs) add to the Cosmic Ray (CR) contribution from a
few keV to several GeV. The geomagnetic field deflects GCR particles depending
on their rigidity and thus the flux of the low rigidity component gets modified.
Intense geomagnetic storms created by the solar activity can affect geomagnetic
shielding and thus affect the interplanetary CR distribution \citep{dorm71}.

Study of the CR interaction and overall radiation environment in the atmosphere
and in space is important for many reasons. The radiation and particles in the
overall CR products affect satellite operations \citep{miro03}, produce
significant background counts in space-based or balloon-borne detectors
\citep{pete75}, affect the crews and passengers in aircrafts \citep{buti11}.

The Monte Carlo (MC) simulation technique has been used in the past to study
CR interactions in the atmosphere. For this, Geant4 \citep{agos03} or FLUKA
\citep{batt07} simulation toolkit is useful. The implementation of microscopic
interactions of the CR particles in event by event basis has been done
by several others \citep{zucc03, heck98, deso05, deso06, pasc14}. We developed
a simulation framework using Geant4 to study CR interactions in the atmosphere
exploiting full 3D implementation of the atmospheric and geomagnetic structure.
We used optimized techniques for efficient simulation and exploited the
advantage of deploying the latest models for the atmosphere, magnetosphere and
interactions in the process.

Though the implementation of this simulation framework allows us to study
detailed radiation environment at various heights in the atmosphere in particular, 
and in space in general, considering several inputs from GCR or solar CR, here in 
the present work, we restricted ourselves to use only major components of the GCR
(viz. H and He) as the input particles. The GCR contribution from other
heavier nuclei has been considered in the He spectrum through a proper scaling
factor in the nucleon fraction \citep{usos17}. In the context of estimating
detector background at certain heights and locations, we estimated the spectral
and directional distribution of several secondaries generated at that particular
locations. However, we also estimate the spectral and directional behaviour at
the satellite height and compare the result with the Alpha Magnetic Spectrometer
(AMS) result \citep{ams00} for benchmark of the simulation.

The estimation and understanding of the background counts generated by the CR
interactions with the detector and surrounding mass distribution is very
important to achieve required sensitivity of the detector, particularly for the
high-energy $\gamma$-ray astrophysics where the source intensity could be low. An
accurate information of the particle and radiation distributions at the payload
environment and proper simulation of the interaction of these particles in the
payload from $4\pi$ solid angle is necessary to develop a proper background
rejection algorithm and evaluate the remaining background. There are some models
describing the radiation environment which can be used for this purpose
\citep{mizu04, lei06}. However, it is more appropriate to generate the desired
distributions as a function of energy and observation direction for a given
location and a given solar condition. This would be a very ambitious project,
however, it is the only way the appropriate corrections can be made.

In the present work, we consider simulation of CR interactions in the atmosphere
with realistic conditions to generate the particle distribution, a potential
contributor to the background counts during measurements of extraterrestrial
radiation in X-rays during our light-weight paradigm in balloon-borne study
(e.g., \cite{chak17, sark17} and references therein). We generally use
scintillation detectors in these experiments, typically working in the energy
range of $\sim$ 15-100 keV. The detector ascends through the atmosphere to
reach near $\sim$ 40-42 km. The estimation of the background counts is important
for these detectors like other high-energy astronomical experiments. The major
contributors to these experiments are the secondary radiation and particles from
the CR interaction with the atmospheric nuclei. The cosmic gamma-rays also can
contribute to the background, depending on the location of the detector. Apart
from the direct contribution of these cosmic primary or secondary particles and
radiation, there can be a significant background from the induced radioactivity
in the detector and surrounding materials due to high energy cosmic-ray
particles, along with some natural radioactivity in the detector material
\citep{pete75}. To understand the background counts in the detector, we
calculate the radiation distribution from the simulation of CR interaction in
the atmosphere. We use this distribution as input to the further simulation to
calculate background counts in the detector with proper mass model of the
deployed payloads using Geant4 simulation toolkit.

In the following section (Sec. \ref{sec:simu}), we describe the general
simulation procedure and considerations of the CR interactions with the
atmosphere and compare the result at satellite heights with AMS measurement
\citep{ams00} to validate the simulation. Section \ref{sec:atbal} gives the
particle distribution scenario at balloon altitude, which is used in the next
Sec. \ref{sec:comp} to evaluate the background counts in a balloon-borne
detector. Finally, we discuss the results from this work in Sec. \ref{sec:conc}.

\section{Simulation Procedure for Atmospheric Interactions}
\label{sec:simu}
For the simulation of CR interactions with the Earth atmosphere we need to
consider the following aspects. (1) An accurate distribution of the atmospheric
molecules inside of which the incident CR interacts to produce the secondary
particle shower. (2) The magnetic field distribution which controls the
trajectories of the charged particles also defines the cutoff rigidity for those
particles. (3) The spectra of the primary CR components at the top of the
atmosphere. (4) The physical processes which describe the interactions of the CR
particles with the atmospheric nuclei and atoms. All these simulation aspects
have been realized by implementation of a full 3D model of the atmosphere and
magnetosphere using Geant4 simulation toolkit with proper distribution of the
primary particles and considering suitable interaction processes in Geant4.

\subsection{Atmospheric model}
\label{ssec:atmos}
The distribution of matter in the atmosphere depends on the location, time and
other conditions, such as, solar activity. This leads to different interaction
probabilities of the particles and radiation varying dynamically with location
and time. We considered the NRLMSISE-00 standard atmospheric model
\citep{pico02} up to 100 km from the Earth surface with proper input parameters
for location, time and solar condition (parameterized by solar 10.7 cm radio
flux and magnetic A$_p$ index) to describe the atmosphere in the simulation. The
location and time used for the simulation was chosen to match with the
experimental conditions we conducted. The solar 10.7 cm flux and A$_p$ index
value were fixed, depending on the time of the experiment, at 100 $\times$
10$^{-22}$ watt/m$^2$/Hz and 5 respectively for the simulation \citep{solind}.
We have limited the atmospheric height till 100 km considering the fact that
most of the high-energy CR particles interact in the atmosphere below this
altitude. This consideration reduces the simulation load because the step size
of the particle tracks are relatively smaller in the atmospheric medium
compared to other regions without any matter distribution. The whole
atmosphere is subdivided into 100 concentric spherical layers of equal
logarithmic altitude (in km). We calculated the temperature, pressure,
total mass density and number density of each of the major molecular components
(e.g. N$_2$, O$_2$, He, Ar, H, N, O) in the atmospheric layers from the existing
model to construct the whole atmosphere.

\subsection{Geomagnetic model}
\label{ssec:geomag}
The magnetic field surrounding the Earth affects both the primary and secondary
CR charged particles and subsequently the generation of neutral particles
through deviation and entrapment resulting in the rigidity cutoff of incoming
charged particles. It also defines the trajectories of all the charged
particles. The field distribution is highly dynamic and strongly coupled to the
solar activity in addition to the relatively slower secular variation in the
geomagnetic field. The magnetic field in the vicinity of the Earth has two
components: (i) the inner magnetic field due to the Earth's magnetism near the
Earth surface which extends up to about 4 Earth radii from Earth surface and
(ii) the external magnetic field which depends on the Interplanetary Magnetic
Field (IMF) and solar conditions. We consider the magnetic field distribution up
to 25 Earth radii. The inner magnetic field used in this simulation is
calculated using the 12th generation IGRF model \citep{theb15} with proper
input parameters. The external magnetic field is calculated using Tsyganenko
Model \citep{tsyg16}. For our simulation purpose we considered solar parameters,
required in the magnetospheric models during the time of 92nd Dignity mission
\citep{chak17} of Indian Centre for Space Physics (ICSP). We will use the
simulation results to compare and understand the data obtained from this
mission. The date and time of interest is: 11th May, 2016 at around 5:35 UT.
During the simulation of the charged particle tracks in Geant4, the
resulting magnetic fields are calculated at the position of each step of the particle
track and the track is recalculated integrating the equation of motion of the
charged particle in the field. We used {\it G4CashKarpRKF45} stepper method
\citep{geant4f} provided in Geant4 for the integration purpose.

\subsection{Primary particle generation}
\label{ssec:part}
In the context of the present work, where we study the interaction of
the cosmogenic particles to produce secondary particles in the atmosphere, we
only consider the GCR. The most abundant components of GCR are the protons (H)
and helium nuclei (He) which constitute about 99\% of the GCR flux. So we
consider these two species of the CR particles along with the Cosmic Diffuse
Gamma-Ray Background (CDGRB) \citep{dean91} as the main inputs of our simulation 
in order to evaluate the cosmogenic background of the X-ray detector. Here
in this work, we are not considering the primary CR electrons because our present
study is confined in the relatively low latitude where the cut-off energy is
high and also the trigger efficiency in the detector under consideration for
electrons is very low.

The primary radiation and neutral particles reach the Earth undeviated and
interact with the atmosphere, while the charged particles are deflected by
the Earth's magnetic field. In the ideal case, we need to
generate an isotropic distribution of particle flux at a great distance where the
flux distortion due to magnetic field is negligible and then calculate the
particle flux at the desired location (say, at the top of the atmosphere). This
is clearly an inefficient process for the simulation. Instead, we follow the
prescription for the backtracing method as considered by \cite{zucc03} and in
PLANETOCOSMICS \citep{deso06}. We produce an isotropic distribution of the
charged particles from a geocentric spherical surface at $500$ km above the Earth
surface and backtracked them to reach an outer surface at $25$ Earth radii in
presence of the magnetic field distribution described in Sec. \ref{ssec:geomag}.
We select only those particles which reach the outer sphere as the allowed GCR
particle tracks and proceed with the simulation for atmospheric interactions.

Based on the direct GCR observations in the very local intersteller medium
by {\it Voyager 1} \citep{ston13} after it crossed the heliopause and using the
PAMELA \citep{adri13} and AMS02 \citep{agui15} data, new model to represent the
very local interstellar spectrum (LIS) has been proposed by \cite{vos15}. In
this work we used the same LIS represented by:
\begin{linenomath*}
\begin{equation}
J_{LIS}(E_k) = N \frac{E_k^{1.12}}{\beta^2} \left(\frac{E_k +
0.67}{1.67}\right)^{-\alpha},
\label{eqn:lis}
\end{equation}
\end{linenomath*}
where, $E_k$ is the kinetic energy of the particles, $N$ - normalization factor,
$\alpha$ - index of the power-law. The modified differential particle flux at 1
AU is given by \citep{herb17}:
\begin{linenomath*}
\begin{equation}
J(E_k, \phi) = J_{LIS}(E_k + \Phi)
\frac{E_k (E_k + 2Mc^2)}{(E_k + \Phi)(E_k+\Phi + 2Mc^2)},
\label{eqn:prim}
\end{equation}
\end{linenomath*}
where, $M$ - particle mass, $c$ - speed of light and $\Phi$ = $(Z/A)\phi$. $Z$
and $A$- atomic and mass number of the particle and $\phi$ - solar modulation
parameter. To produce the primary H flux we considered $A$ = $2.70 \times 10^3$
$particles\, s^{-1} m^{-2} sr^{-1} GeV^{-1}$ and $\alpha$ = $3.93$. The solar
modulation parameter $\phi$ is fixed at 524 MV \citep{usos17} for the time
considered during the simulation for comparison with the balloon-borne
experiment while $\phi$ = 555 MV was considered for the comparison with AMS
result in Sec. \ref{ssec:atsat}. For the He spectrum we use $\alpha$ = $3.89$,
which is required to fit the He spectrum given in \cite{biss16}. The
normalization factor is modified using the nucleon fraction for He particles
with an additional factor of 0.3 which also includes the contribution from other
heavier species in GCR \citep{usos17}.

\begin{figure}
  \centering
  \noindent\includegraphics[width=0.48\textwidth]{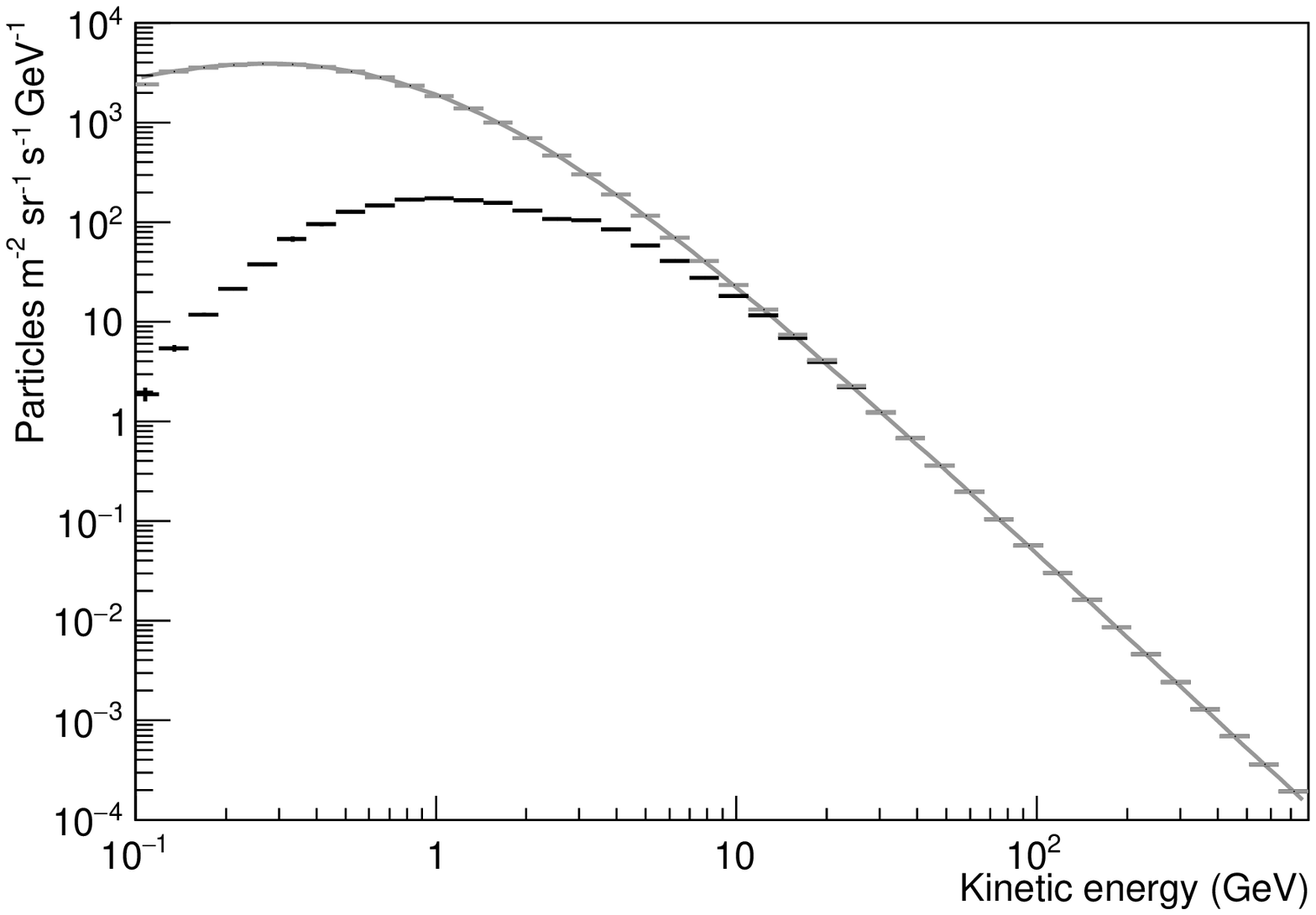}
  \noindent\includegraphics[width=0.48\textwidth]{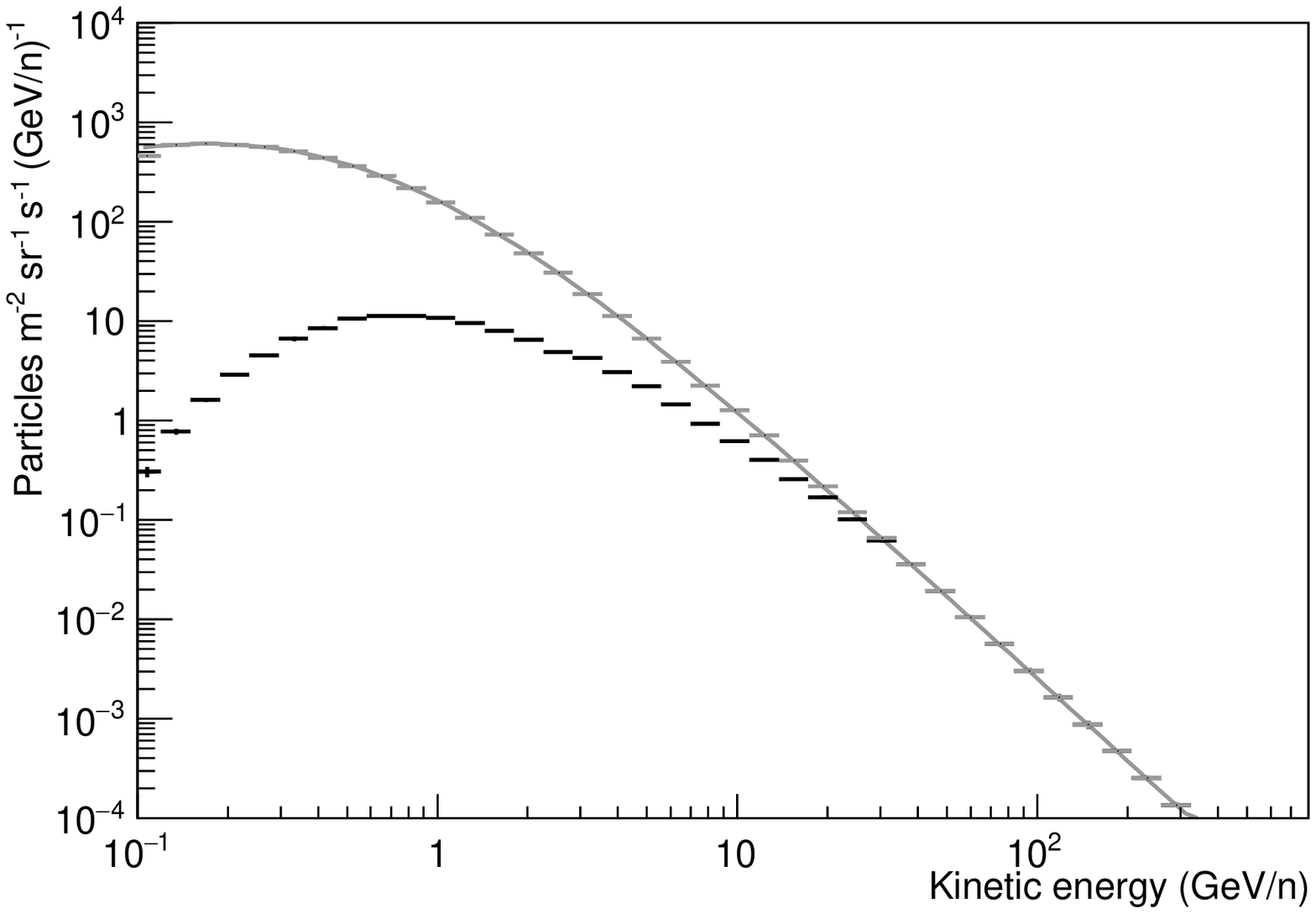}
  \caption{Primary particle flux considered in the simulation: (left) H flux;
  (right) He flux. Gray data points represent original primary flux
  generated according to Eqn. \ref{eqn:prim}. Black points represent the same
  fluxes after the geomagnetic cutoff (see text for details).}
  \label{fig:prim}
\end{figure}

Apart from the solar modulation, a more prominent cause of modulation of the
primary flux comes from the rigidity cutoff due to geomagnetic field when the
charged particles approach the Earth. This modulation is achieved in the
simulation through the incorporation of the accurate geomagnetic field
configuration. The generated primary flux of the H and He and modulation due to
the geomagnetic cutoff is shown in Fig. \ref{fig:prim}. This is the average
flux over the geomagnetic latitude region of $0-57.3^{\circ}$. Here the
rigidity cutoff of the charged particles is manifested from the simulation
inherently due to the interaction of the individual particle tracks with the
defined geomagnetic field and there is no need to separately supply the rigidity
cutoff value to modify Eqn. \ref{eqn:prim}.

Considering the geomagnetic cutoff effect at low energy and decreased particle
flux at high energy due to power-law distribution, we restricted the primary
generation in the energy range of $0.1-800.0$ GeV/n. The particles in this
energy range carry $\sim 98\%$ of the total energy of CR particles allowing us
to include most of the energy transport interactions in our simulation
\citep{zucc02}. To maintain the statistical significance of counts in the whole
energy range, we subdivided the whole energy range in 5 sub-ranges (0.1-0.3,
0.3-3.0, 3.0-30.0, 30.0-170.0 and 170.0-800.0 GeV/n). We simulated a total of
$3.5\times10^5$ particles both for H and He and calculated the particle
abundance to be simulated in each energy range, keeping in mind the number of
secondaries produced in the energy range. This is done so that the statistical
significance of the secondary counts in each range remains comparable.

Because of our particular interest in relatively low energy photons, we also
simulated CDGRB photons in the energy range of 10 keV - 0.1 GeV. For this
purpose, we used the primary photon spectrum observed by {\it Swift BAT}
\citep{ajel08} for energies $\leq$ 1 MeV and for higher energies up to 0.1 GeV
we considered {\it COMPTEL} observation \citep{weid99}:
\begin{linenomath*}
\begin{equation}
J(E) = \begin{cases}
\frac{N_1}{\left(\frac{E}{E_b}\right)^{\alpha_1} +
\left(\frac{E}{E_b}\right)^{\alpha_2}}, & \text{for E $\leq$ 1 MeV,}\\
N_2 \left(\frac{E}{5\, MeV}\right)^{-\alpha_3}, & \text{for E $>$ 1 MeV,}
\end{cases}
\label{eqn:phot}
\end{equation}
\end{linenomath*}
where, $N_1 = 1.015 \times 10^9$, $N_2 = 1.12 \times 10^3$ $photons\, s^{-1}
m^{-2} sr^{-1} GeV^{-1}$, $\alpha_1 = 1.32$, $\alpha_2 = 2.88$, $\alpha_3 =
2.2$ and $E_b = 29.99$ keV. The whole energy range is divided into 4 sub-ranges
(10-100, 100-10$^3$, 10$^3$-10$^4$ and 10$^4$-10$^5$ keV) and simulated 10$^5$
photons in each energy sub-range to maintain statistical significance of
counts.

\subsection{Physical interaction models}
\label{ssec:int}
To simulate interaction of high energy particles with the atmosphere we
need an optimal interaction model. We used the reference physics list: QGSP
physics list with Binary Cascade model {\it QGSP\_BIC\_HP} \citep{geant4},
provided in Geant4 which is recommended for cosmic-ray applications and optimally
covers the interaction processes relevant here. In this current simulation
procedure, the relevant energy interval is 10 keV - 800 GeV. The hadronic
interactions considered in the physics list are recommended to be suitable
at least up to 10 TeV, while the low energy limit mentioned to be valid from 0.
To handle the electromagnetic part this model this physics list use the
``standard" Geant4 electromagnetic physics which is valid from 100 eV to 100
TeV. This physics list includes the high precision neutron interaction model to
handle the elastic and inelastic scattering, capture and fission more accurately
from 20 MeV down to thermal neutrons.

\subsection{Validation of simulation using AMS data}
\label{ssec:atsat}
\begin{figure}
  \centering
  \noindent\includegraphics[width=0.48\textwidth]{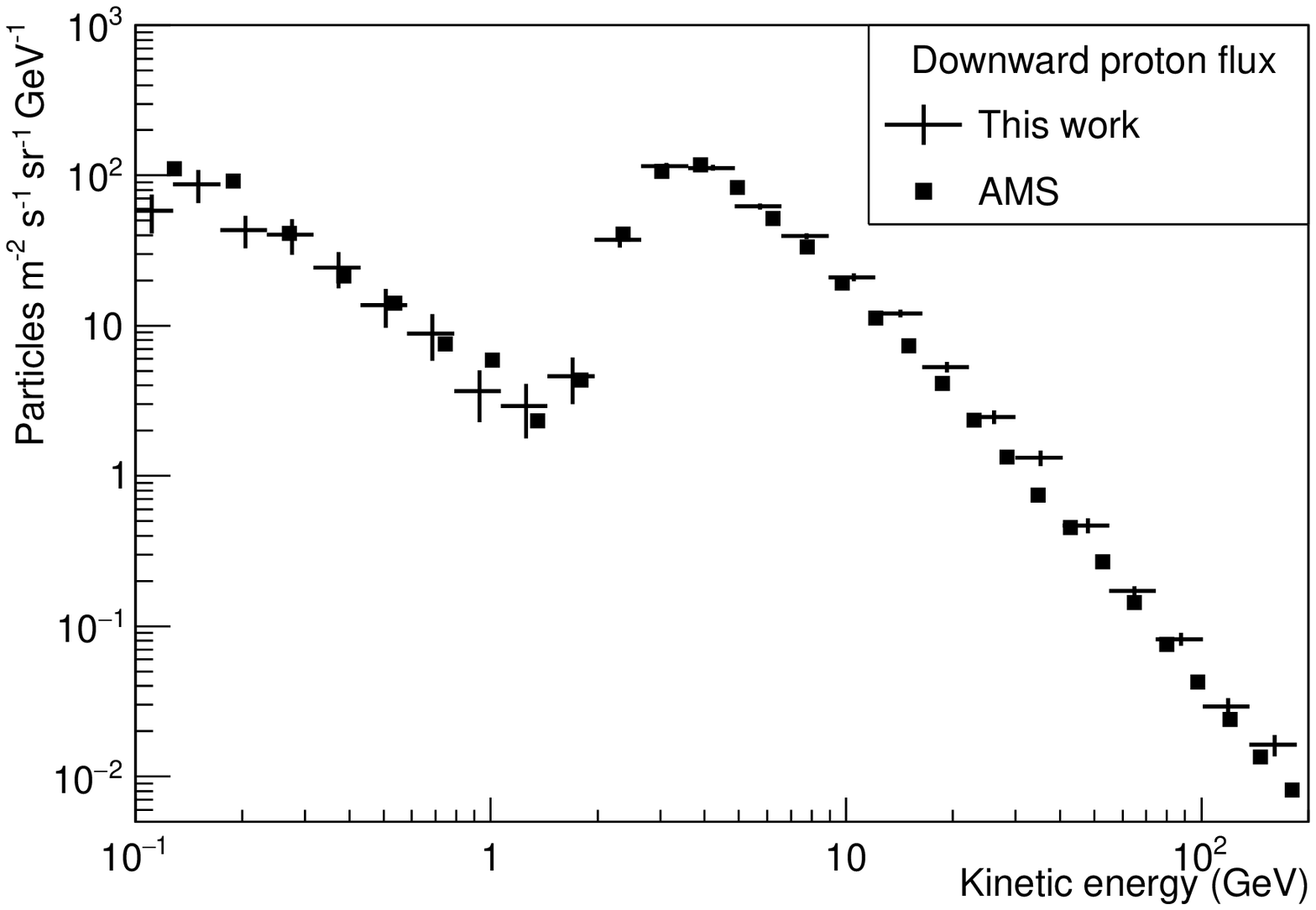}
  \noindent\includegraphics[width=0.48\textwidth]{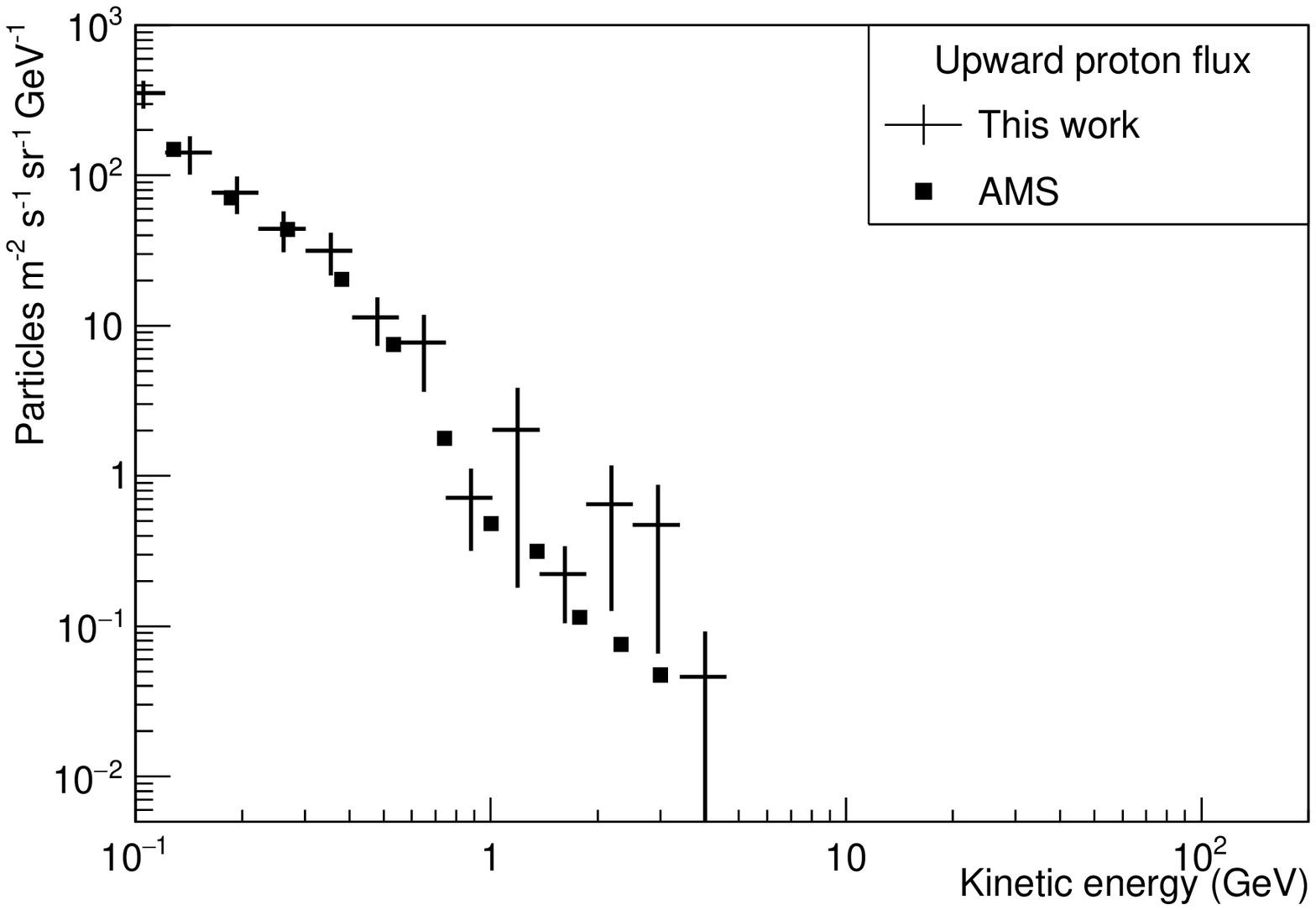}
  \caption{Downward/upward proton flux at the satellite height of 400 km due
  to H and He particle interactions in the latitude range of $40.1^{\circ} <
  \theta_M < 45.8^{\circ}$ and their comparison with AMS data.} 
  \label{fig:prot400}
\end{figure}

To validate the simulation procedure by comparing the results with the AMS
proton flux measurement \citep{ams00}, we calculated the total proton flux
(from primary CR and secondary generated) at the satellite altitude of $\sim$
400 km and within $32^{\circ}$ of AMS z-direction (here we consider the
zenith and nadir directions respectively for downward and upward particles).
The simulation input parameters for the atmospheric and magnetic field models
were adjusted according to the time of the AMS observation used here for
comparison. We considered the calculations in the corrected geomagnetic latitude
($\theta_M$, as mentioned in \cite{ams00}) range of $40.1^{\circ} < \theta_M <
45.8^{\circ}$. In Fig. \ref{fig:prot400} we show the downward and upward proton
fluxes calculated from the simulation. The corresponding AMS measured fluxes are
 presented in the same plot for comparison. We also conducted a goodness-of-fit
test to quantify the agreement of the results by using Pearson's $\chi^2$ test.
For the downward protons this test yielded a $\chi^2$/NDF value 16.44/18 with
probability factor 0.56 and for the upward proton we found the respective values
to be 12.93/11 and 0.30. Thus the calculated results are in general agreement with
the AMS results.

\section{Particles at Balloon Height}
\label{sec:atbal}
To understand the background counts in a detector at the balloon altitude we
calculated the flux distribution of different CR generated secondary particles
such as: protons, neutrons, electrons, positrons, muons (+/-) and photons at an
altitude of $\sim$ 30 km (atmospheric depth $\sim$ 11.52 g/cm$^2$). Depending on
the mass distribution model of the payloads, these particles may interact and
produce counts in the detectors.

\begin{figure}
  \centering
  \noindent\includegraphics[width=0.45\textwidth]{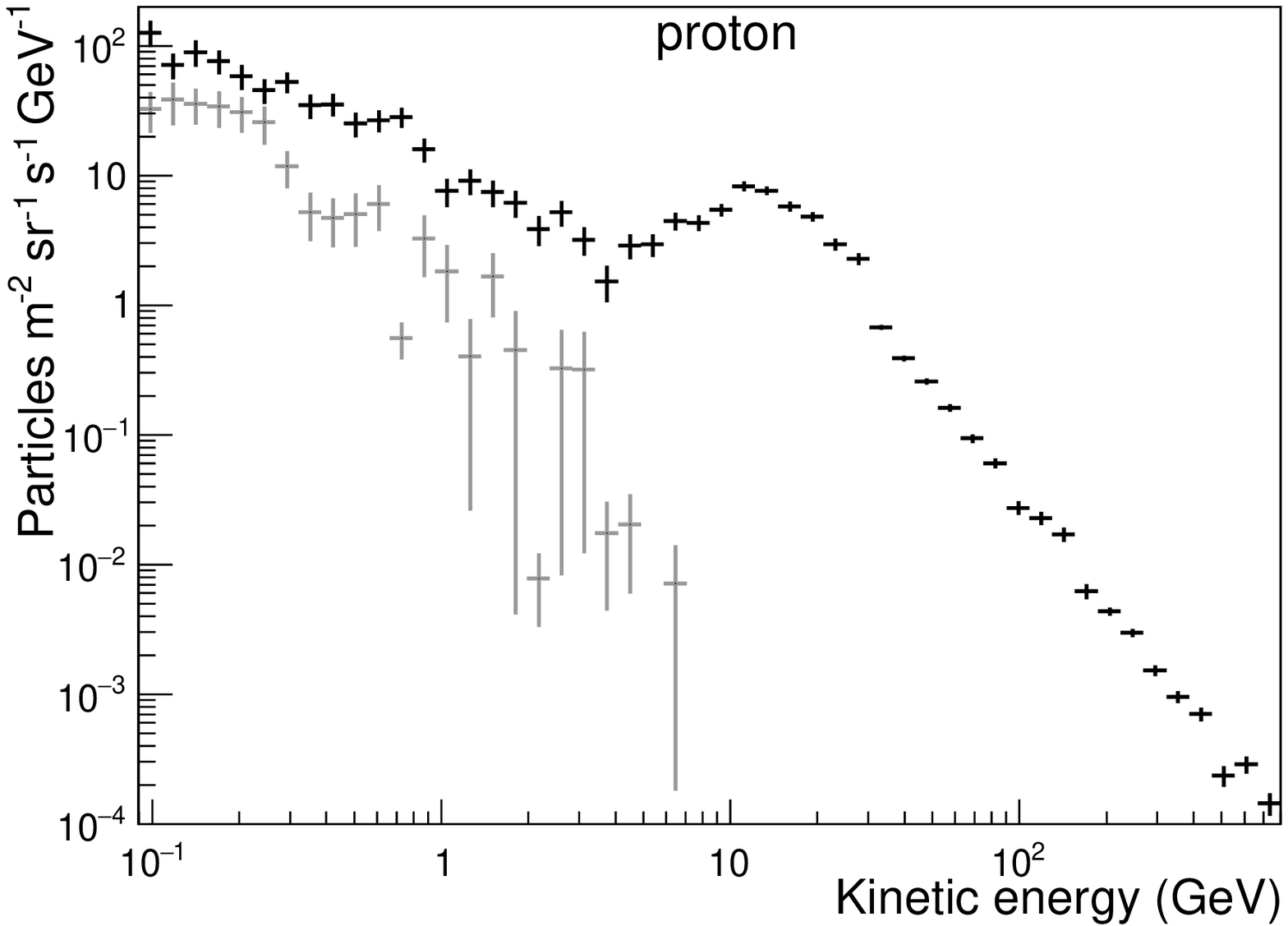} \\
  \noindent\includegraphics[width=0.45\textwidth]{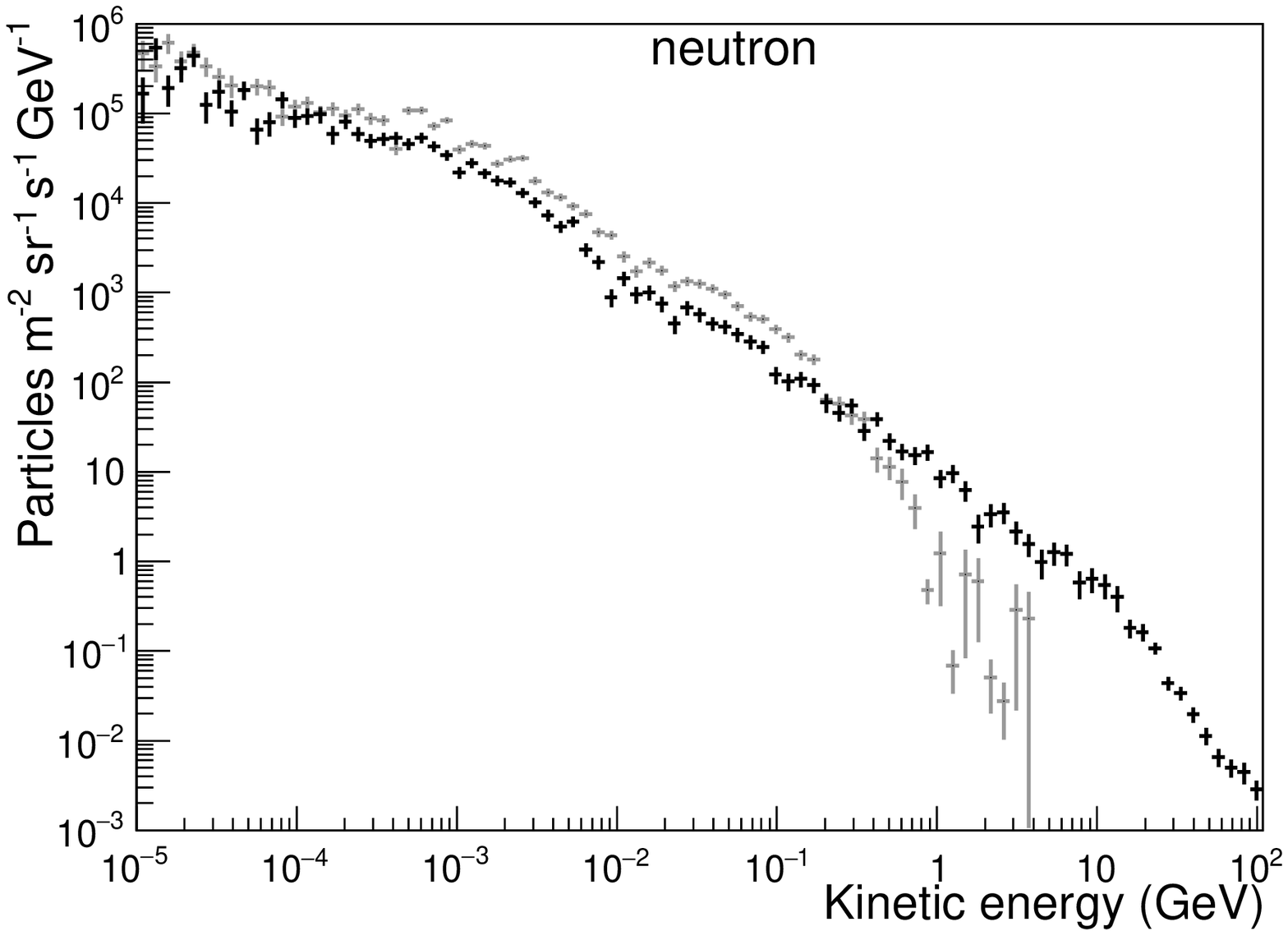}
  \noindent\includegraphics[width=0.45\textwidth]{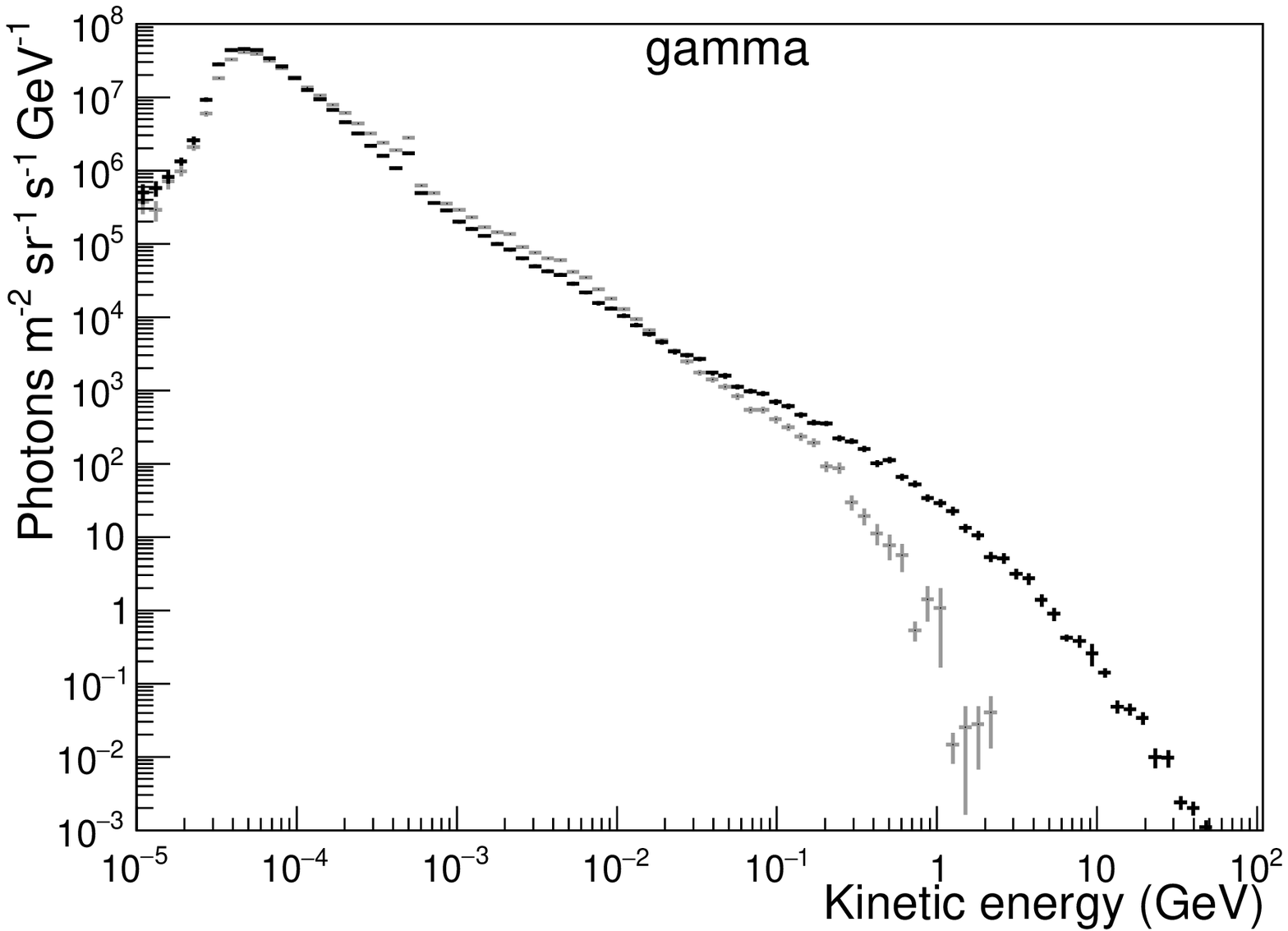}
  \noindent\includegraphics[width=0.45\textwidth]{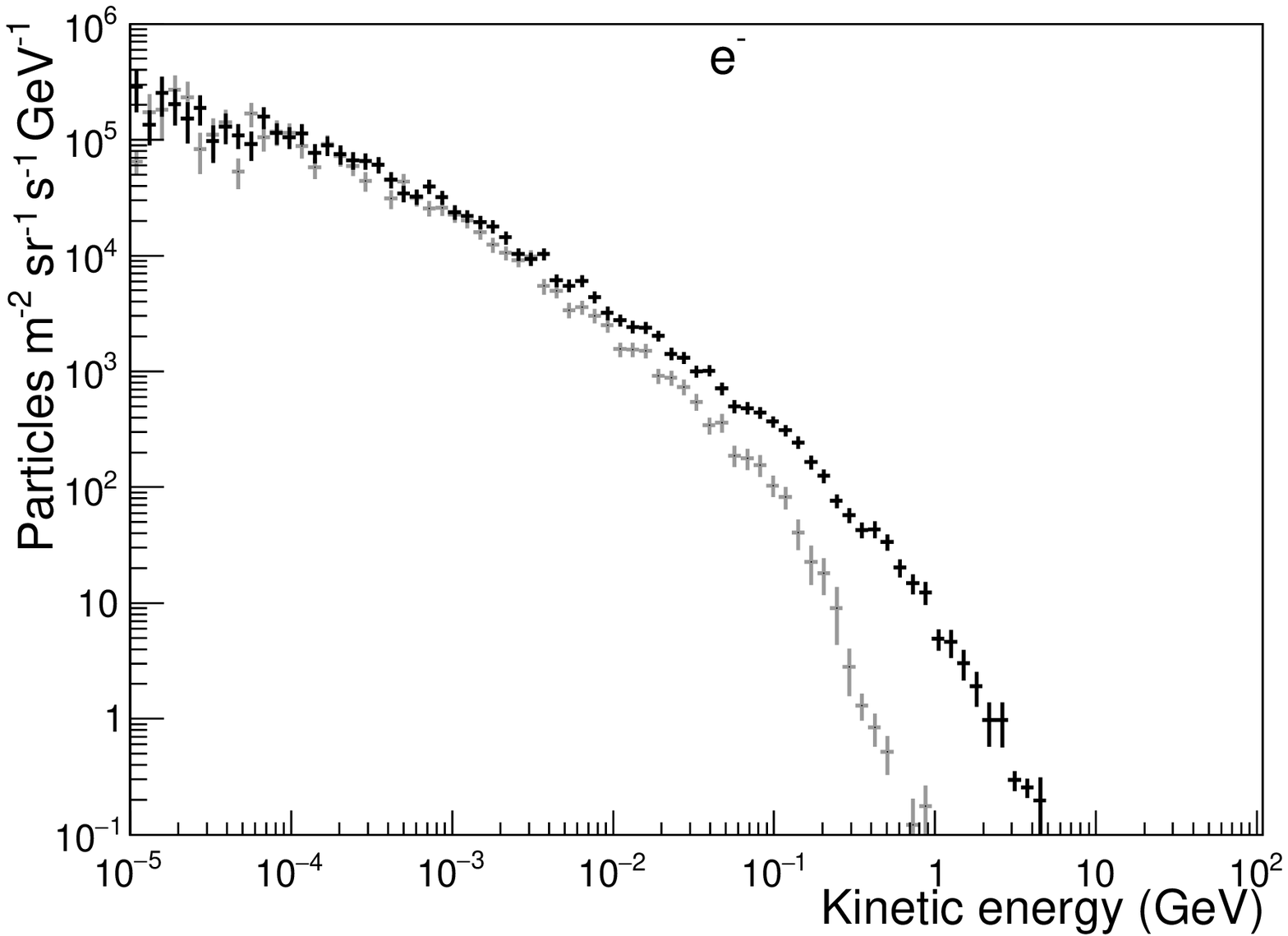}
  \noindent\includegraphics[width=0.45\textwidth]{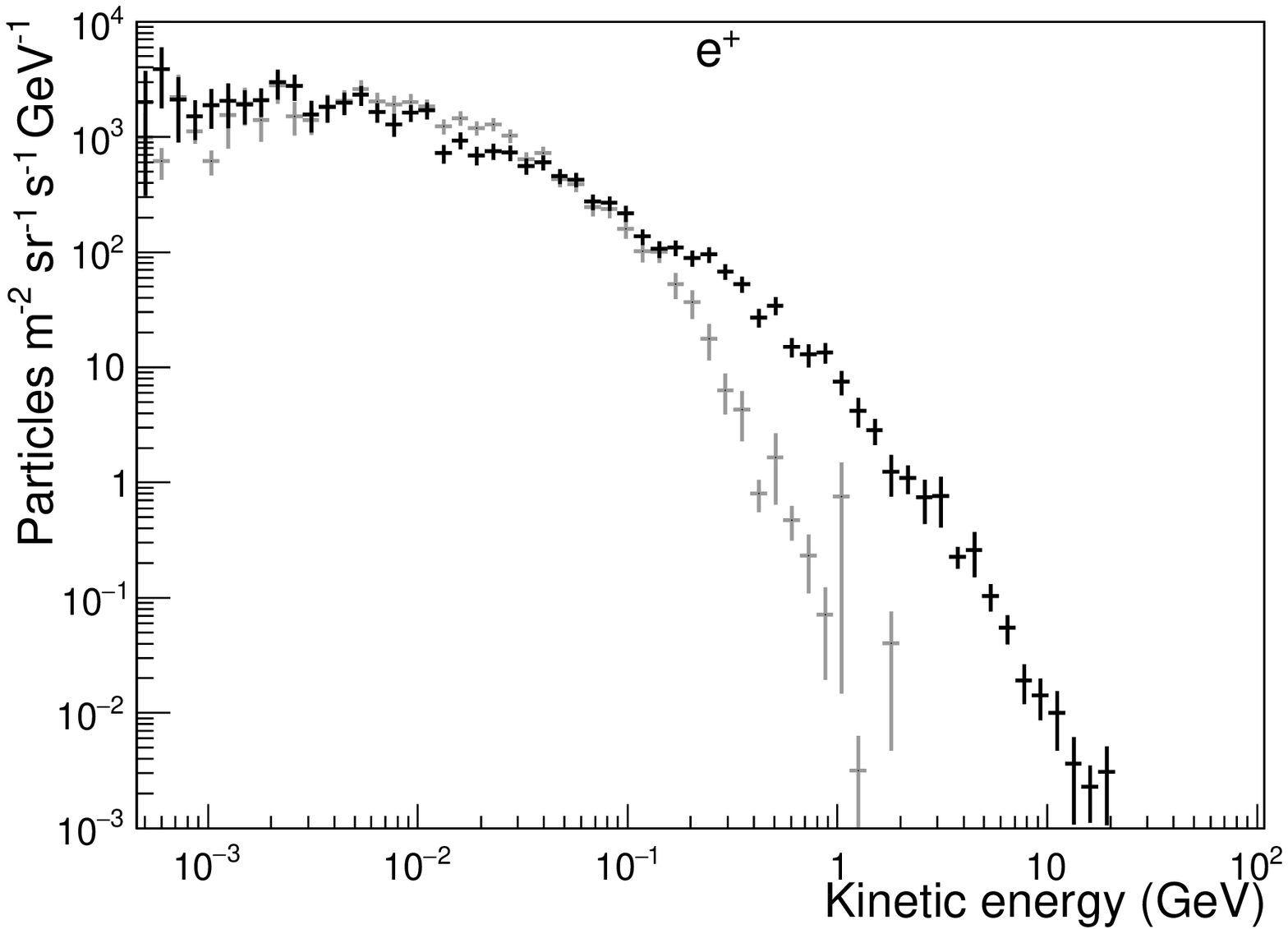}
  \noindent\includegraphics[width=0.45\textwidth]{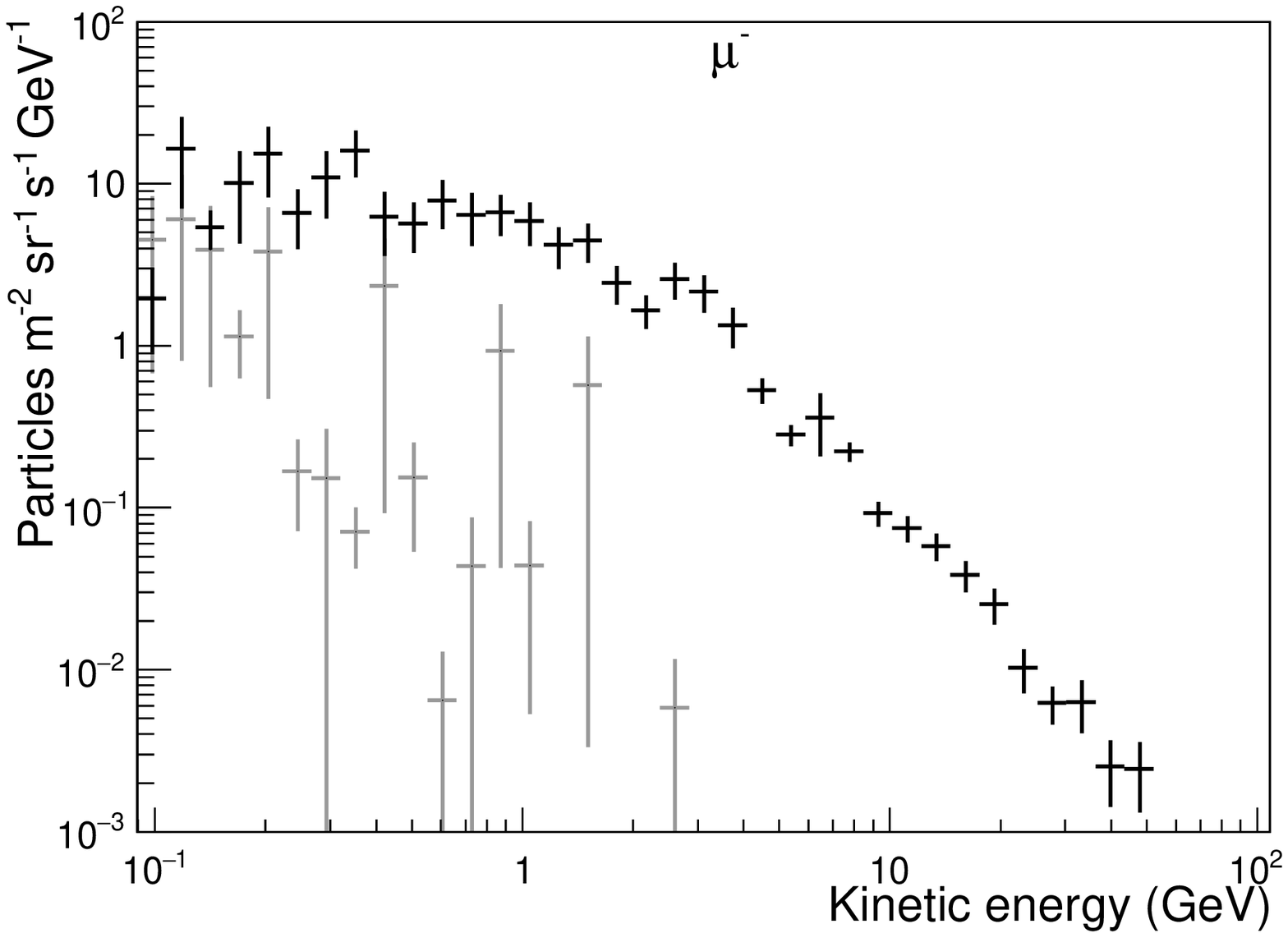}
  \noindent\includegraphics[width=0.45\textwidth]{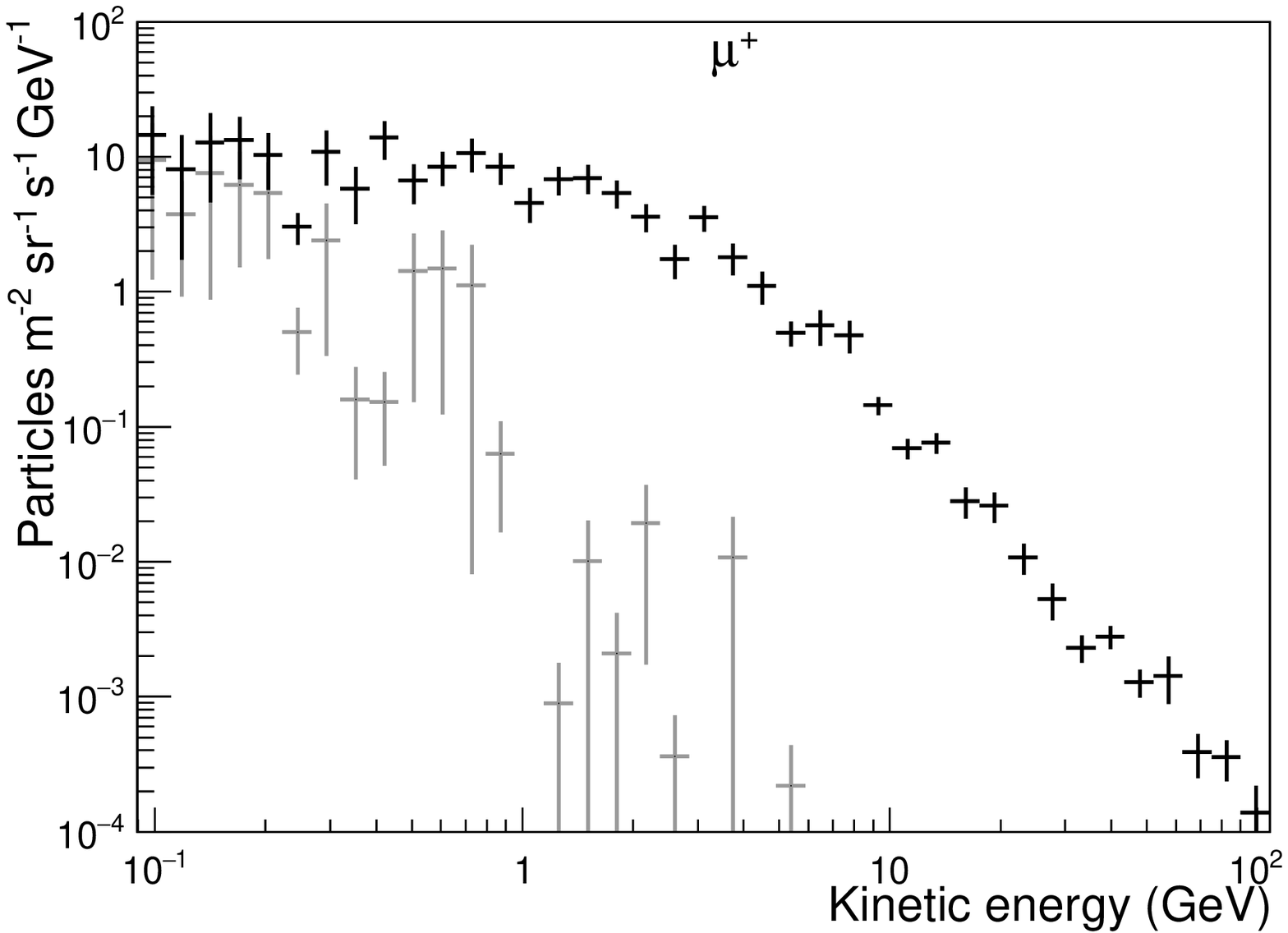}
  \caption{Downward (black)/upward (gray) particle flux for proton, neutron,
  photon, $e^-$, $e^+$, $\mu^-$ and $\mu^+$ at balloon altitude of 30 km ($\sim$
  11.52 g/cm$^2$) and in the latitude range of $11.5^{\circ} < \theta_M <
  17.2^{\circ}$ due to cosmic ray H, He and CDGRB photon interactions
  considering whole 90$^{\circ}$ angle from zenith/nadir.}
  \label{fig:part30}
\end{figure}

The spectral distribution of the particles at the balloon altitude and at 
geomagnetic latitude of $11.5^{\circ} < \theta_M < 17.2^{\circ}$ are shown in
Fig. \ref{fig:part30}. The choice of this particular latitude is to calculate
and compare the background of a balloon borne experiment near the tropic of cancer
which is discussed in more detail in Sec. \ref{sec:comp}. Here we considered to
calculate the particle distributions within whole 90$^{\circ}$ angle from zenith
and nadir for downward and upward particles respectively. This has been done to
include the effect of directional distribution of the particles interacting with
the payload mass.

The secondary particles at this height is much more abundant than the
primaries as compared to those at the satellite height. Though, the downward and
upward proton fluxes differ with diminishing contribution of the upward
protons, the upward and downward neutron and photon fluxes are comparable
at least at the lower energy domain. Similar behavior, to some extent, can be
seen for the e$^-$/e$^+$ which (generated from CR interactions in the
atmosphere) are scarce at the satellite height. However for the muon component,
the downward counterparts are more prominent.

\section{Comparison with Observed Data}
\label{sec:comp}
We used energetic and directional distribution of different particles at the
balloon altitude obtained from the simulation described in Sec. \ref{sec:atbal},
as the input for the simulation to estimate the counts due to these particles in
a balloon-borne detector. This detector was deployed in an experiment designed
to measure extraterrestrial radiation from astronomical sources. In this case,
the secondary and primary CR particles at balloon altitude interacts with the
payload mass, as well as with the active detector volume to generate detector
counts which contribute to the detector background. Estimation of the background
and its reduction is very crucial for these detectors as they deal with
relatively low source counts.

\subsection{Experiment overview}
\label{ssec:expt}
Indian Centre for Space Physics has been measuring CR induced radiations in
the atmosphere and high-energy X-rays from astrophysical sources for over a
decade through its low-cost, near-space program with rubber or polystyrene
meteorological balloons. Here, light-weight payloads are used without any
pointing device or heavy shielding which make the data reduction tedious in the
absence of an accurate model of background radiation. The details of the program
can be found in \cite{chak14, chak15, chak17}. In this paper, we concentrate on
the results obtained in Dignity 92 mission. Our scientific payload comprised of
a phoswich scintillator detector which consists of a combination of 3 mm
thick NaI(Tl) and 25 mm thick CsI(Na) crystal discs of 116 mm diameter mounted
on a Photo-Multiplier Tube (PMT) housed in an aluminum casing. The detector is
fitted with a collimator (made of 0.5 mm thick lead) to reduce the
off-axis external background counts. The detector works in the energy range of
$\sim$ 15-100 keV and counts in the detector are triggered by the energy
depositions in the scintillator crystals. However, the detector is operated in
anticoincidence mode to record only exclusive full energy deposition by an event
(photon) in the NaI crystal. The detector counts are recorded in the on-board
data storage system during the mission flight with the help of suitable readout
and data processing system. There are other ancillary instruments to accomplish
the mission. More detail about the detector and the payload may be found in
\cite{chak17, sark18}. 

In the mission under consideration, the payload was launched from Muluk, West
Bengal, India (latitude: 23.64$^{\circ}$N; longitude: 87.71$^{\circ}$E;
geomagnetic latitude: 14.5$^{\circ}$N) on May 11, 2016 at 03:42 UT. The payload
went up to $\sim$ 41 km ($\sim$ 2.03 g/cm$^2$), however, for the sake of
general comparison and understanding of the experimental data using our
simulated result, we considered the experimental data at 30 km (11.52
g/cm$^2$) during the ascend of the payload.

\begin{figure}
  \centering
  \noindent\includegraphics[width=0.50\textwidth]{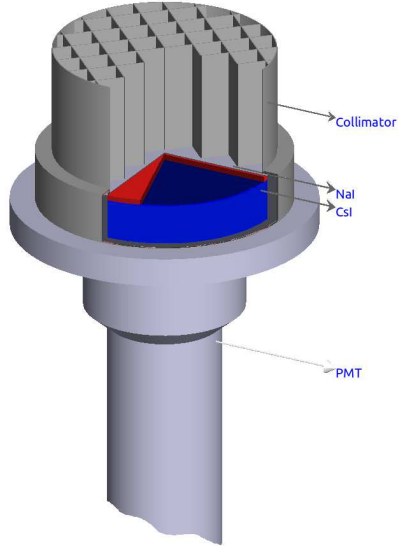}
  \caption{Mass model of the phoswich detector used for the background
  simulation.}
  \label{fig:phos}
\end{figure}

We performed a detailed simulation of the detector to calculate the background
counts due to the CR generated particles and radiations using the Geant4
simulation framework. A proper mass model of the payload is crucial for these
types of simulation as the materials used in the payload construction may produce
significant radiation counts which add to the detector background. However, due
to the unique payload design of the mission under consideration where the
payload casing is mainly made up of styrofoam, the mass distribution of the
payload frame and other ancillary instruments can be neglected. The electronic
circuits and other ancillary instruments are very small in volume and
weight and are not considered in the simulation. The mass model of the detector
considered for the simulation is shown in Fig. \ref{fig:phos}.

\subsection{Detector background calculation}
\label{ssec:bkg}
There are several contributors to the background counts for the astronomical
detectors on board balloons or satellite platforms. The contributions may be
from the external sources such as: CR generated particles, trapped particles, CDGRB
radiation etc. or due to the radiations internal to the detector, such as, induced
radioactivity from the activated isotopes, induced spallation radiation in the
active detector, natural radioactivity from the detector or surrounding materials etc.
\citep{pete75}.

\begin{figure}
  \centering
  \noindent\includegraphics[width=0.6\textwidth]{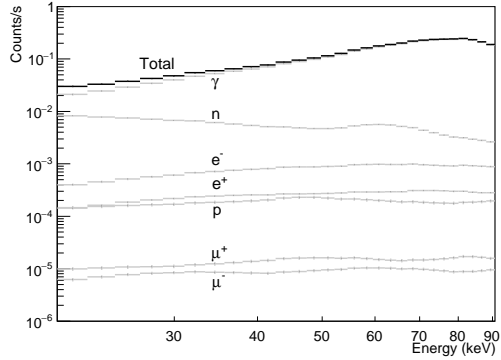}
  \caption{Contribution to the detector counts due to different cosmogenic
  radiation and particles. The total CR contribution is shown in black while the
  individual components are plotted in gray.}
  \label{fig:crbkg}
\end{figure}

Here, in this work we have explicitly calculated the contribution to the detector
background due to cosmogenic particles and radiation. For this purpose we used
the information of the simulated production of secondary CR particles along with
the CDGRB contribution as the input to the simulation of the detector
background. We used the spectral information of the secondary particles in the
energy range which is able to generate significant number of triggers in the
detector. We also took into account, the zenith angle distribution of the
particles and radiation during the generation of the input particles for the
simulation. However, due to the axial symmetry of the detector along the zenith
direction we did not consider the azimuthal distribution of the particle
generation. Digitization of the energy deposition in the NaI crystal was
thoroughly done by incorporating the anti-coincidence effects with CsI crystal
and response function of the detector obtained during its calibration to produce
the background counts in the detector \citep{sark19}. The contribution of
several cosmogenic particles and radiation produced by the CR interaction in the
atmosphere along with the CDGRB photons, to the detector background counts is
shown in Fig. \ref{fig:crbkg}. Our result shows that the major contributors to
the detector background are photons ($\sim$ 95\%) and neutrons ($\sim$ 4\%) of
the secondary CR.

\begin{figure}
  \centering
  \noindent\includegraphics[width=0.6\textwidth]{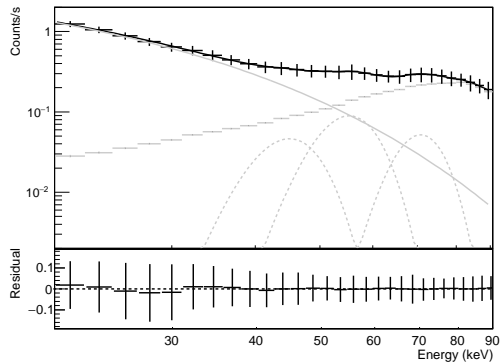}
  \caption{Components of different background contributors to the total
  background of the detector at 30 km altitude. The total detected background
  is given by black points; external CR contribution calculated from the
  simulation given by gray points, low-energy exponential form of the induced
  internal radioactivity given by gray solid line with three other Gaussian
  (dashed gray lines); the total calculated background is shown by black solid
  line along the detector data points. The lower panel shows the residual to the
  fitting.}
  \label{fig:bkgcomp}
\end{figure}

However, the external background from the CR origin comprises only a part of
the total background of the detector, in the energy range under consideration, as it
is evident from Fig. \ref{fig:bkgcomp}. The lower energy part is dominated by
the internal background of the detector by the induced radioactivity due to
activation by capture of neutrons in the detector or/and shielding material or
by spallation radioactivity due to breakup of high Z target nucleus in the
scintillator crystal. This cosmogenic induced background depends on the position
of the payload in the atmosphere \citep{sark19}, as the secondary cosmic ray
flux varies with the altitude and location in the atmosphere. For instance, the
observed count rate by the detector near 20 keV (where internal background is
dominating) can be approximated by the exponential form: $exp(0.116 + 0.013
\times D_{atm})$, where $D_{atm}$ is the atmospheric depth at different altitude (in
the range 20 - 40 km or 55.77-2.38 g/cm$^2$).

The induced background spectrum produced in the detector depends on several
factors: the CR particle flux inducing the radioactivity in the detector and
surrounding materials, production rate of several radioactive isotopes, decay of
the radioactive isotopes in the allowable observation time and efficiency of
detection of the decay counts in the detector \citep{fish72}. The final
background spectrum due to these induced activity composed of several
decay lines and continua from beta decay. Here, in this work we have not
calculated  the induced background from the radioactive products in detail.
Instead, to explain the total background counts in the detector, we have
empirically fitted the induced background part using four components. The lower
energy part has been treated with an exponential cutoff of the form:
\begin{linenomath*}
\begin{equation}
\frac{dF}{dE} = \frac{F_c}{E_f} \exp\left(-\frac{E}{E_f}\right)
\end{equation}
\end{linenomath*}
as described by \cite{fish72}. Here, $F_c$ represents the rate of detected
counts and $E_f$ is the e-folding energy. The value of $F_c$ depends on the
incident CR and hence on the location of the detector. We also needed to use
three other Gaussian lines due to radiation from the induced or natural
radioactive isotopes $^{210}$Pb at $\sim$ 44 keV, $^{125}$I at 67.2 keV
\citep{adhi17} and $^{127}$I at 57 keV \citep{pete75}. The spread of these
Gaussian lines was fixed to the resolution of the detector at the corresponding
energies but their normalization factors were free for the fit. For the
background spectrum of the detector at the location under consideration, the fit
gives $F_c$ = 77.79 $counts/s$, $E_f$ = 13.29 keV and three normalization
factors as 0.046, 0.089 and 0.052 $counts/s$. The internal background along with
the external CR background explains the detector background satisfactorily. The
observed background spectrum along with the external CR contribution calculated
from the simulation and fitted background components are shown in Fig.
\ref{fig:bkgcomp}. The lower panel of the Figure shows the residual values. The
total background count rate at 30 km altitude in the energy range of 20-90 keV is
12.28 counts/s. This counting rate varies with the atmospheric depth ($D_{atm}$)
roughly as the exponential form: $exp(2.34 + 0.0148 \times D_{atm})$. Thus,
for example, the expected count rate at the height of 40 km ($D_{atm}$ = 2.38
g/cm$^2$) is approximately 10.8 counts/s in the same energy range.

\section{Conclusions}
\label{sec:conc}
Along with several other important implications, the interaction of Galactic
cosmic rays in the atmosphere plays a crucial role in the astronomical
experiments, particularly for the balloon borne detectors. The background counts
produced in the detectors by these CR generated particles limit the
detectibility of the astrophysical sources and demand a thorough study. Here in
this work, in the context of a light-weight balloon-borne experiment to study
extraterrestrial radiation sources, we studied the CR generated particles in the
atmosphere and their effects on the detector. In this course of work, we
developed a simulation framework using Geant4 toolkit to study the interaction
of the extraterrestrial radiation and particles with Earth atmosphere in
presence of the geomagnetic field. We validated the simulation results at
satellite height using the AMS result. Then, we calculated the spectral and
angular distribution of various secondary particles at the balloon height due to
the interaction of H and He (effectively including the contribution from
heavier nuclei) which are the most abundant species of cosmic rays.
Although, we considered only the interaction of the GCR with the atmosphere,
this simulation framework can be extended to study other interactions, such as, the
effects of SEP in our atmosphere.

We used the results produced by the CR interaction with the atmosphere to
estimate the background counts in a balloon borne astronomical detector.
However, the external CRs produce only a part of the detector background in its
operating energy range and we also need to consider the internal background
induced by the CRs or energetic particles of other origins. Here, we have
treated the problem of internal background only empirically which can be
calculated more extensively using detailed analytical \citep{dyer71} and
simulation \citep{adhi17} procedures.

\section*{Acknowledgments}
The authors would like to thank the ICSP balloon team members, namely,
Mr. D. Bhowmick, Mr. A Bhattacharya, Mr. S. Midya, Mr. H. Roy, Mr. R. C. Das 
and Mr. U. Sardar for their valuable supports in various forms during the
mission operations and data collection. This work been done under partial
financial support from the Science and Engineering Research Board (SERB,
Department of Science and Technology, Government of India) project no.
EMR/2016/003870. We also thank the Higher Education department for a Grant-In-Aid 
which allowed us to carry out the research activities at ICSP.
All the data shown in this work are available from the authors.


\end{document}